\documentclass[modern]{aastex61}

\usepackage{graphicx}% Including figure files
\usepackage{amsmath}% Advanced maths commands
\usepackage{amssymb}% Extra maths symbols
\usepackage{enumitem}

\newcommand\mybar{\kern1pt\rule[-\dp\strutbox]{.8pt}{\baselineskip}\kern1pt}

\setlist[itemize]{noitemsep, topsep=0pt, leftmargin=*}

\shorttitle{Flares from Space Debris in LSST}
\shortauthors{Loeb}

\begin{document}

\title{Flares from Space Debris in LSST Images}

\author{Abraham Loeb}
\affiliation{Astronomy Department, Harvard University, 60 Garden
  St., Cambridge, MA 02138, USA}

\begin{abstract}
Owing to the exceptional sensitivity of the Vera C. Rubin Observatory,
we predict that its upcoming LSST images will be contaminated by
numerous flares from centimeter-scale space debris in Low Earth Orbits
(LEO). Millisecond-duration flares from these LEO objects are expected
to produce detectable image streaks of a few arcseconds with AB magnitudes
brighter than 14.

\end{abstract}

\section{Introduction}

The European Space Agency (ESA)
reports\footnote{\url{https://www.esa.int/Space_Safety/Space_Debris/Space_debris_by_the_numbers}}
that as of December 6, 2023, the space debris in orbit around Earth
includes $1.3\times 10^8$ objects in the size range of 0.1-1~cm, $\sim
10^6$ objects between 1-10~cm and $3.65\times 10^4$ objects larger
than 10~cm. A subset of these objects is in Low Earth
Orbits (LEO) with a semi-major axis below an altitude of $2\times
10^3~{\rm km}$. In this Note, we examine the
implications of this LEO debris for the upcoming Legacy Survey of Space \&
Time (LSST) of the Vera C. Rubin Observatory in
Chile~\citep{Necht,Ivezic,Bosch,Esteves,Hernandez,Schwamb}.

\section{Method}

We define the average albedo of an object of radius $R$ and distance
$d$ which is illuminated on one hemisphere by sunlight by the following fraction
of light reflected from it,
\begin{equation}
A\left({\pi R^2\over 2\pi d^2}\right)~.
\label{eq:0}
\end{equation}
Hereafter, we encapsulate the unknown surface properties of the
object as well as the orientation and geometry of the source relative
to the Sun and the observer in the value of $A$, for which we adopt a
fiducial value of 0.1. Given the average solar illumination of
$f=1.4\times 10^6~{\rm erg~s^{-1}}$ at a characteristic photon frequency of
$\nu \sim 6\times 10^{14}~{\rm Hz}$, we calculate the AB
magnitude of the object to be,
\begin{equation}
{\rm AB}=-2.5\log_{10}\left[{(df/d\nu)\over {\rm erg~s^{-1}~cm^{-2}~Hz^{-1}}}\right]-48.6=16.2 - 5 \log_{10}\left[\left({R_{0}\over d_8}\right)\sqrt{A_{-1}}\right] ,
\label{eq:1}
\end{equation}
where $A_{-1}=(A/0.1)$, $R_{0}=(R/1~{\rm cm})$ and $d_8=(d/10^8~{\rm cm})$.

\section{Results}

Data from the Zwicky Transient Facility (ZTF) shows that the sunlight
glints from known LEO satellites generate flashes of
duration~$10^{-3\pm0.5}~{\rm s}$ (see Figure 4 in~\citet{Karpov}, and
also~\citet{Hank,Guy2,Guy1}).  Given a typical orbital speed of $\sim 8~{\rm
    km~s^{-1}}$, these flash durations translate to streaks of angular
  length $\sim (10^{-3}~{\rm s}\times 8~{\rm km~s^{-1}})/10^3d_8~{\rm
    km})= 1.7 d_8^{-1}$ arcseconds. The light from the flares is
  therefore expected to spread across no more than a few arcseconds,
  independently of the LSST exposure time which is 4 orders of
  magnitude longer.

For a few millisecond flares, LSST at the Rubin Observatory~\citep{Ivezic}
is expected to have a brightness sensitivity of ${\rm AB}\sim
14$~(given the survey's $\sim 25$-magnitude limit for 30 second
exposures; see Figure 5 in~\citet{Karpov}). According to
equation~(\ref{eq:1}), this implies that LSST will detect all space
debris objects in LEO that satisfy,
\begin{equation}
\left({R_{0}\over d_8}\right)\sqrt{A_{-1}}\gtrsim 3 .
\label{eq:2}
\end{equation}
This includes all LEO objects with a radius $R\gtrsim (3/A_{-1})~{\rm
  cm}$ at a distance of $d\sim 10^3~{\rm km}$.

\section{Discussion}

Assuming that merely a tenth of the $\sim 10^6$ space debris objects
in the size range of 1-10~cm reported by ESA satisfy the condition in
equation~({\ref{eq:2}), we find that LSST images will be contaminated
  by flares from~$\sim 10^5$ objects that repeat every LEO orbital
  time of 1.5-2 hours (and are most visible during the evening and
  morning twilight). This number exceeds by an order of magnitude the
  number of large satellites currently in orbit around Earth.  Out of
  the entire debris population, only $3.515\times 10^4$ objects
  are regularly tracked and catalogued by Space Surveillance
  Networks.\footnote{\url{https://www.esa.int/Space_Safety/Space_Debris/Space_debris_by_the_numbers}}

So far, the LSST team contemplated a novel strategy to mitigate the
impact of large commercial satellite constellations in
LEO~\citep{Hu}. However, the above numbers suggest that image contamination
by untracked space debris might pose a bigger challenge.

\bigskip
\bigskip
\section*{Acknowledgements}

This work was supported in part by Harvard's {\it Black Hole
  Initiative}, which is funded by grants from JFT and GBMF.  

\bigskip
\bigskip
\bigskip

\bibliographystyle{aasjournal}
\bibliography{t}
\label{lastpage}
\end{document}